\documentclass{article}
\pdfoutput=1

\usepackage{arxiv}
\usepackage[utf8]{inputenc} 
\usepackage[T1]{fontenc}    
 \usepackage{libertine,libertinust1math} 

\usepackage{hyperref}       
\usepackage{url}            
\usepackage{booktabs}       
\usepackage[normalem]{ulem}
\useunder{\uline}{\ul}{}

\usepackage{amsmath}
\usepackage{amsfonts}       
\usepackage{enumitem}
\setlist[itemize]{align=parleft,left=0pt..1em}
\setlist[enumerate]{align=parleft,left=0pt..1.5em,label=\large\protect\textcircled{\small\arabic*}}
\usepackage[font=scriptsize]{caption}
\usepackage{nicefrac}       
\usepackage[final, babel]{microtype}      
\usepackage{graphicx,epstopdf}
\usepackage{doi}
\usepackage{inconsolata}
\usepackage{xcolor}
\usepackage{listings}

\usepackage{hyperref} 
\hypersetup{
  pdftex,
  final,
  colorlinks=true,
  linkcolor=blue,
  filecolor=magenta,
  urlcolor=cyan,
  breaklinks = true,
  pdfstartview = FitV,
  linktocpage = false,
}

\title{SD-BLS}

\author{
  Denis Roio \\
  Dyne.org \\
    \And
  Rebecca Selvaggini \\
  UniTN \\
    \And
  Gabriele Bellini \\
  UniMI \\
    \And
  Andrea D'Intino \\
  Forkbomb B.V.
}

\renewcommand{\headeright}{Info @ Dyne.org}
\renewcommand{\undertitle}{Privacy Preserving Selective Disclosure of Verifiable Credentials with Unlinkable Threshold Revocation}
\renewcommand{\shorttitle}{\textit{SD-BLS}}

\hypersetup{
  pdftitle={SD-BLS: Privacy Preserving Selective Disclosure of Verifiable Credentials with Unlinkable Threshold Revocation},
  pdfsubject={cs.CR},
  pdfauthor={Denis Roio, Rebecca Selvaggini, Gabriele Bellini, Andrea D'Intino},
  pdfkeywords={Privacy, Selective disclosure, BLS signatures, Digital credentials, Revocation}
}

\begin{document}

\twocolumn[
  \begin{@twocolumnfalse}

\maketitle

\begin{abstract}
 Ensuring privacy and protection from issuer corruption in digital identity systems is crucial. We propose a method for selective disclosure and privacy-preserving revocation of digital credentials using second-order Elliptic Curves and Boneh-Lynn-Shacham (BLS) signatures. We make holders able to present proofs of possession of selected credentials without disclosing them, and we protect their presentations from replay attacks. Revocations may be distributed among multiple revocation issuers using publicly verifiable secret sharing (PVSS) and activated only by configurable consensus, ensuring robust protection against issuer corruption. Our system’s unique design enables fast revocation checks allowing the use of large revocation lists.
\end{abstract}

\vspace{2cm}

 \end{@twocolumnfalse}
]
\section{Introduction}
Digital identity systems implement credential issuance and
presentation mechanisms so that a person (holder) can voluntarily
disclose his or her own acquired skills, professed attributes, or
completed accomplishments. Credentials are signed by issuer
authorities and encapsulated within various forms of digital proofs to
be held in digital wallets, empowering individuals to reveal only
chosen details to designated recipients, to limit data exposure and
permit a user-controlled release of information.

Such systems are known as selective disclosure and they enhance users
privacy by allowing data minimization when proving credentials.

\subsection{State of the art}

Selective disclosures are being used by nation states across the world
in their next generation identity systems, for instance EIDAS2.0 in
Europe where the European Digital Identity Wallet Architecture and
Reference Framework\cite{eudi-arf} mandates the use of
SD-JWT\cite{sd-jwt} and mDOC \cite{mdoc}. SD-JWT adopts for its
cryptography Hash-Based Message Authentication Codes (HMAC) to
generate proofs: such presentations can be traced and recent critical
feedback on the EUDI ARF\cite{troncoso} details this problem.

In North America the efforts concentrate on the adoption of the
BBS+ algorithm\cite{bbs+} leveraging its Zero Knowledge Proof
properties and applied to W3C Verifiable Credentials\cite{w3c-vc} to
obtain an higher degree of privacy by making every disclosure
unlinkable.

\subsection{Threats considered}

The different choices in data formats in these two approaches is
irrelevant in relation with cryptography, any choice between JSON Web
Tokens or W3C Verifiable Credentials does not impact the privacy
level. But the cryptography adopted determines the adequacy of a
solution to face three important threats that can render an algorithm
unsuitable to be used in real world situations.

\paragraph{Linkability}

The EUDI-ARF standard dictates that credentials \textit{issued} to a
holder, can be \textit{presented} (in the form of a verifiable
presentation) to a relying party in order to have one or more
attributes verified. Every verifiable presentation includes one or
more HMAC(s), formatted in SD-JWT: the HMACs are identical each time a
verifiable presentation is produced from a certain credential. This
makes possible for colluding relying parties, or to malicious actors,
to trace a holder's identity by collecting, exchanging and confronting
verifiable presentations (linkability). This threat appears to be well
mitigated by BBS+ through its Zero Knowledge Proof implementation.

\paragraph{Privacy breach of revocation lists}

We believe that anonymous revocations are a \textit{condicio sine qua
  non} to guarantee a sufficient level of privacy in digital
identities and credentials. There is no privacy-preserving revocation
system designed, either in EUDI-ARF, or in W3C-VC and BBS+. In case
the choice of strategy for revocation is left open to developers, the
risk for major privacy breaches may occur, for example with the
adoption of public status lists \cite{crlcomparison}. The hypotetical
use of a certificate status lists (CRL) presents issues related
primarily to privacy \cite{CRL}, because sensitive information about
holders leaks from the list. This problem is partially mitigated by
expiration dates, in cases where credentials can be short-lived
(typically less important credentials), but not applicable with
digital identification documents such as ID, driving license, passport
and social security numbers, which typically have longer or no
expiration time. Future plans for the national standards we are
observing include the adoption of ``Bitstring'' status lists
\cite{status-lists} which may grant a degree of privacy. In SD-BLS we
design a privacy-preserving revocation mechanism to remove the leak of
holder's information and delegate the governance of revocations to a
quorum of multiple revocation issuers which may be different from the
credential issuer.


\paragraph{Revocation issuer corruption}

If the choice of interactive revocation is left to a single issuer,
one may unilaterally choose to revoke credentials, without being
subject to revision or having to seek consensus with a quorum of
issuers. This situation leads to security issues in case Issuers are
corrupted and make a weaponized use of digital revocations to
persecute engaged individuals. Such a condition becomes a real concern
for journalists or activists living under dictatorial regimes that may
arbitrarily revoke their credentials, or even ID cards and
passports. Similarly, a security breach of an issuer service, would
result in similar threats. We mitigate this risk by introducing the
possibility for threshold issuance of revocation keys and by
separating the responsibility of revocation issuance and credential
issuance.

\section{Overview}

\subsection{Feature Comparison}

\begin{table}[]
\centering
\renewcommand{\arraystretch}{2}
\begin{tabular}{ccccc}
\hline
       & \textbf{UP} & \textbf{UR} & \textbf{TR} & \textbf{URG} \\ \hline
SD-BLS & no          & yes         & yes         & yes \\ \hline
SD-JWT & no          & wip         & no          & no  \\ \hline
BBS+   & yes         & wip         & no          & yes \\ \hline
\end{tabular}
\vspace*{5mm}
\caption{feature comparison}
\label{tab:features}
\end{table}

We briefly round up on feature differences between the named selective
disclosure cryptographic schemes, as shown in table
\ref{tab:features}, mainly distinguishing between four fundamental
features:
\begin{itemize}
\item UP: Unlinkable Presentation
\item UR: Unlinkable Revocation
\item TR: Threshold Revocation
\item URG: Unregroupability
\end{itemize}

Where \emph{wip} is mentioned, it means work in progress on adoption of
bitstring status lists for unlinkable revocations.

The ``unregroupability'' feature refers to the fact that different
presentations of different claims cannot be linked to each other
(regrouped) as presented by the same holder, even when they are signed
by the same issuer.

\subsection{Key contributions}
The cryptographic scheme described in this paper, named
\textit{SD-BLS} for brevity, implements all the properties of the
SD-JWT scheme and proposes a novel cryptographic approach to similar
data structures. Furthermore SD-BLS proposes novel anonymous
cryptographic revocation flow for verifiable credentials, that aims to
solve governance issues posed by status and revocation lists.

\paragraph{Selective disclosure}
Similarly to the SD-JWT and mDOC formats, SD-BLS produces an array of
claims: the elements of the array are individually signed by the
issuer. In SD-BLS the signature(s) replace the HMAC and still enable
the holder to selectively disclose only certain signed credentials,
and produce a \textit{proof of possession} that minimizes private
information given to verifiers.

\paragraph{Anonymous cryptographic revocation}
SD-BLS proposes a novel approach to credential revocation: the data
published by the revocation issuer will produce cryptographic material
that contains no information about the credential holders. The
cryptographic revocation material allows anyone to verify if an SD-BLS
proof produced by a credential holder has been revoked. The
unlinkability, and thus anonymity of the cryptographic revocation,
allows the revocation issuer to share revocations in public and allows
anyone to verify if credentials have been revoked.

\paragraph{Multi-stakeholder governance of revocations}
SD-BLS allows to introduce a new trusted party to the issuance phase:
the revocation dealer. The dealer doesn't need to know the content of
the credential or the identity of an holder: its role is that of
producing the revocation signature and distributing its secret key to
a configurable range of revocation issuers. Later on a configurable
quorum of issuers may reconstruct the revocation secret key to revoke
a credential. This way the decision on a revocation is not delegated
to a single peer, but to a multi-stakeholder governance that protects
the holder from issuer corruption.

\paragraph{Fast revocation checks}
SD-BLS shows good benchmark results on revocation checks: the main
operation a verifier needs to do is an additional $\mathbb{G}_2$
multiplication. The cost of this operation grows linearly with the
number of published revocations.

\subsection{Applications}
In this section we present some applications and use cases for digital
identity and credentials, that could benefit from using the SD-BLS
scheme.

\paragraph{Digital identity}
The focus of the EUDI-ARF specifications is identity documents: it
defines mechanisms and data structures to issue a Personal
Identification (PID) as, for instance, with digital driving
licenses. Similarly, the US government is experimenting with W3C-VC
and mDOC for cross-states interoperable driving licenses. The SD-BLS
data format is similar to SD-JWT and mDOC, offering selective
disclosure and anonymous revocation.

\paragraph{Academic credentials}
Diploma and academic credentials are among the core offerings of EBSI
as well as a primary research target of the W3C VC working group.

\paragraph{KYC/AML}
We are unaware of standardization efforts for interoperable
credentials in the fields of "Know Your Customer" (KYC) and Anti
Money-Laundering (AML) certifications. We are aware of solution
providers experimenting with W3C-VC for AML applications and believe
SD-BLS can greatly improve the governance of credential revocation,
which is a critical component for this use case.

\paragraph{Generic \textit{light} credentials}
As the digital identity and verifiable credential technologies are
maturing, they are being considered for usage in less privacy
concerning applications, such subscriptions and membership and
fidelity cards.

\paragraph{Verifiable credentials on Blockchain}
SD-BLS can be used with blockchain-based smart-contracts to activate
certain functions:
\begin{itemize}
    \item A proof of possession can be published and be peer-verified
      on-chain in its private form without disclosing the credential
      contents.
    \item A smart-contract may verify if one or more holder's
      credentials match the requirement needed to process a
      transaction.
    \item Issuers can publish their cryptographic revocation lists on
      chain, allowing smart-contracts to verify the status of a
      credential.
    \item The issuer's public keys can also be published on-chain,
      although this does not represent a novelty.
\end{itemize}

\section{Implementation}

In this section we will provide a detailed description of the
algorithm we propose for selective disclosure and unlinkable
revocation using BLS signatures.

\subsection{Notations and assumptions}

We will adopt the following notations:
\begin{itemize}

\item $\mathbb{F}_p$ is the prime finite field with $p$ elements
  (i.e. of prime order $p$);

\item $E$ denotes the (additive) group of points of the curve
  BLS12-381 \cite{bls381-12} which can be described with the
  Weierstrass form $y^2=x^3 + 16$;

\item $E_T$ represents instead the group of points of the twisted
  curve of BLS12-381, with embedding degree $k=12$. The order of
  this group is the same of that of $E$;

\end{itemize}

We also require the notion of a cryptographic pairing. \cite{weilpair}

For the purpose of our protocol we will consider the \emph{Miller
pairing} $e: E_T \times E\to \mathbb{G}_T$, where $\mathbb{G}_T\subset
\mathbb{F}_{p^{12}}$ is the subgroup containing the $n$-th roots of
unity, and $n$ is the order of the groups $E$ and $E_T$.\\ For
completeness we also recall the main properties of the map:

\begin{itemize}

\item [i.] \emph{Bilinearity}, i.e. given $P_1,Q_1\in E_T$
  and $P_2,Q_2\in E$, we have
  \begin{align*}
    e(P_2,P_1+Q_1) = e(P_2,P_1)\cdot e(P_2,Q_1)   \\
    e(P_2+Q_2,P_1) = e(P_2,P_1)\cdot e(Q_2,P_1)
  \end{align*}

\item[ii.] \emph{Non-degeneracy}, meaning that for all
  $g_1\in E_T, g_2\in E$, $e(g_2,g_1)\ne
  1_{\mathbb{G}_T}$, the identity element of the group
  $\mathbb{G}_T$;

\item[iii.] \emph{ Efficiency}, so that the map $e$ is easy to
  compute;

\item[iv. ] $E_T \ne E$, and moreover, that
  there exist no efficient homomorphism between $E_T$ and
  $E$.

\end{itemize}

\subsection{Issuance} \label{issuance}

As for other well known algorithms BLS signing works following three
main steps:
\begin{itemize}

\item Key Generation phase.\\ For an issuer who wants to sign a
  credential $m$, a secret key $sk$ is a random number chosen
  uniformly in $\mathbb{F}_n$, where $n$ is the order of the groups
  $\mathbb{G}_1, \mathbb{G}_2, \mathbb{G}_T$. The corresponding public
  key $pk$ is the element $sk\cdot G_2\in E_T$;

\item Signing phase.\\ The credential $m$ is first hashed into the
  point $U\in E$; the related signature is then given by $\sigma =
  sk\cdot U$;

\item Verification phase.\\ For an other user that wants to verify the
  authenticity and the integrity of the message $m$, it needs to

  \begin{itemize}

  \item [1.] parse $m, pk$ and $\sigma$

  \item [2.] hash the message $m$ into the point $U$ and then
    check if the following identity holds,

    \[
    e(pk,U) = e(G_2,\sigma)
    \]

  \end{itemize}
If verification passes it means that $\sigma$ is a valid signature for
$m$.
\end{itemize}

BLS signatures also support aggregation: it is possible to aggregate a
collection of multiple signatures $\sigma_i$ (each one related to a
different message $m_i$) into a singular new object $\sigma$, that can
be validated using the respective public keys $pk_i$ in a suitable
way.

Since $\sigma_i\in G_1 \forall i$, the algorithm has an homomorphic
property. We exploit this property to add a revocation signature into
the signed credential.

The issuer create a credential as follow: given a claim $m$, it
generate a new secret revocation key $rev$ with public key $r$. Let
$\mathcal{H}$ be a cryptographic hash function, we compute:
\begin{equation*}
    H = \mathcal{H}(m : r)
\end{equation*}
then the issuer proceeds to sign with both its private key, and the
revocation key generated above:
\begin{equation*}\label{rev_agg}
    \begin{split}
        r &= rev \cdot G_2 \\
        \sigma_{rev} &= sign(rev, H : r)\\
        \sigma &= sign(sk, H : r) + \sigma_{rev}\\
   \end{split}
\end{equation*}
The set of all the signed claims will be:
\begin{equation*}
   \mathcal{C} = \big\{ \{H, r, \sigma, m \} : m\in claims  \big\}
\end{equation*}

At the end of this phase the holder is sent the \textit{signed claims}
to be stored in a private wallet, while the issuer stores a list of
\textit{revocations} as tuples formed by $\{H,rev\}$ into a private
database that can be used later to issue revocations.

\subsection{Presentation} \label{presentation}

Any presentation of a SD-BLS credential can simply omit the message
$m$ to separate disclosure from verification as required by the
specific context of a proof of possession and to satisfy privacy
preserving design patterns for data minimization.

\paragraph{Basic Proof}

A credential holder can choose any set of signed claims to present,
and selectively disclose them into what we name ``Basic Proof''
presentation.

The holder can present a basic proof of possession $\{ H, r, \sigma
\}$ for each claim requested, or the complete set $\{ H, r, \sigma, m
\}$ for claims whose content must be disclosed.

However, Both forms of presentation of a basic proof are vulnerable to
replay attacks: any verifier receiving such presentations can reuse
them to impersonate the holder in any other session.

\paragraph{One Time Proof}

To prevent replay attacks we exploit once again the homomorphic
property of BLS signatures to add a sessions signature into the
credential signature, then we add separate fields containing a signed
timestamp and the public key of the session signature.

We name this presentation ``One Time Proof''.

For each presentation the holder generates a key pair $(sk_t, pk_t)$
and a string $t$ containing session information, i.e. an expiration
date or a pointer to the intended recipient. Then we compute:
\begin{equation*}
    \begin{split}
        \sigma' &= \sigma + sign(sk_t, H : r) \\
        \sigma_{t} &= sign(sk_t, H : r : \sigma' : t : pk_t)
    \end{split}
\end{equation*}
And obtain a presentation composed as follows:
\begin{equation*}
    \{H, r, \sigma', t, pk_t,  \sigma_t\}
\end{equation*}

This schema is secure because it is impossible to reconstruct the
holder information $\{H,r,\sigma\}$ necessary for a reply attack;
indeed, given $\sigma'$, an attacker that wants to retrieve $\sigma$
should be able to reconstruct the signature $\sigma_t$,
but this is not possible without the knowledge of the secret key $sk_t$.

Furthermore the three attributes $t,\sigma_t, pk_t$ cannot be modified
or removed because $pk_t$ is necessary for the verification of
signature $\sigma'$.  If an attacker tries to create a new valid
presentation with freshly generated $sk_t', pk_t', t'$ and
$\sigma_t'$, then the signature $\sigma'' = \sigma' + sign(sk_t',
H:r)$ can be verified using the public key $pk_t + pk_t'$.  However
this public key does not verify the signature $\sigma_t'$, and
$\sigma_t'$ can not be updated using the homomorphic property since it
is the signature over the old $\sigma'$.  Thus any tampering of the
presentation will lead to an invalid credential.

\subsection{Verification}

Credential verification is made by checking the presented issuer's
signature and revocation status of each claim. In order to verify the
signature the credential issuer's public key must be added to the
revocation public key.

\paragraph{Basic proof}

We consider the basic credential proof as a collection of tables of
the following form:
\begin{equation*}
    c = \{H, r, \sigma \}
\end{equation*}
where $\sigma$ and $r$ are respectively the signature of the string
$H$ and the revocation public key, optionally the claim value $m$ can
be included.

We can check the validity of the presented claim computing the key:
\begin{equation*}
    pk = A.pk + r
\end{equation*}
and verify the bls signature $\sigma$.

As proof of correctness consider that the signature $\sigma$ is given
by:
\begin{equation*}
\begin{split}
    \sigma &= sign(A.sk, H:r) + sign(rev, H:r) \\
     &= A.sk U + rev U
\end{split}
\end{equation*}
where $U$ is the mapping of the string $H$ in the group $G_1$.\\
Recalling the verification formula, it holds that:
\begin{equation*}
\begin{split}
 e(pk, U) &= e(A.skG_2 + revG_2, U) \\
 &= e(A.skG_2, U) \cdot e(revG_2, U) \\
 &= e(G_2, A.sk U) \cdot e(G_2, rev U) \\
 &= e(G_2, A.sk U + rev U) \\
 &= e(G_2, \sigma) \\
\end{split}
\end{equation*}
where each equality holds for the bilinearity of the Miller loop.

When the presentation contains the value $m$ it should also be checked
that:
\begin{equation*}
    H = \mathcal{H}(m : r).
\end{equation*}

\paragraph{One Time Proof}

If the verifier receives the credential locked to the current session
(with replay attack protection) as $\{H, r, \sigma', t, pk_t,
\sigma_t\}$ then it proceeds to
\begin{itemize}
\item verifiy the information in $t$ (e.g. timestamp)
\item verify $\sigma_t$ with public key $pk_t$
\item verify $\sigma'$ with the public key $A.pk + r + pk_t$
	\end{itemize}

In case the verifier is presented with a one time disclosure that
includes $m$ then it needs also to check its hashed value $H =
\mathcal{H}(m : r)$.

This concludes the first verification phase. If the given presentation
is valid, then the verifier should proceed to check the revocation
status of the credential.

\subsection{Revocation}

To control if a revocation has been emitted for any credential being
verified, we update the revocation list from the issuer. A revocation
list can be publicly distributed since revocation keys do not provide
any information on the identity of holders.

Given an element $c = \{H, r, \sigma \}$ of the credential
presentation, and a series of $rev$ elements found in any revocation
list, we can verify a revocation by checking if the revocation public
key presented by the holder matches:
\begin{equation*}
    r = G_2 \cdot rev
\end{equation*}
This step should follow the signature verification and is sufficient
to guarantee the revocation status.

If a dishonest holder provides a wrong $r$, then the verification of
the signature will fail, and the given credential should be considered
invalid.

If it happens that the signature verification is successful and that
the revocation key does not match, one can conclude that the
credential is not revoked.

Note that, in the case the verifier received a credential in the ``One
Time Proof'' form, the check for the revocation status does not
change.

\subsection{Threshold Revocation}

In order to split responsibilities over interactive revocation we
introduce a threshold over the revocation key, plus we split the
functionality of \textit{credential issuance} from that of
\textit{revocation issuance}, now operated by different peers.

This is implemented via an interactive process facilitated by a third
party trustee, a revocation dealer, who should be:
\begin{itemize}
    \item never entitled to publish revocations
    \item connected to credential issuers to complete any credential
      signature
    \item never informed about the identity of credential holders
    \item regularly connected to revocation issuers to distribute
      shares
\end{itemize}

Such a revocation dealer will be in contact with \textit{credential
  issuers} for the signature of credentials: it will create the $rev$
revocation key while concealing it from them. The dealer then proceeds
creating the $\sigma_{rev} = sign(rev, H : r)$ revocation signature
and the $r = rev \cdot G_2$ public key, which will be communicated to
the credential issuer to be aggregated into the signed credential (see
section \eqref{rev_agg}).

The dealer will then proceed to split the secret revocation key into
shares using a public verifiable secret sharing (PVSS \cite{pvss})
implementation and distribute these shares to all revocation
issuers. In order to issue a revocation, a configurable quorum of
peers among the revocation issuers will need to reconstruct the secret
revocation key and publish it.

This process separates responsibilities between the credential issuer
and the revocation issuers, delegating to the revocation issuers the
possibility to revoke a credential interactively through a collective
process.

The collection of shares can be done asynchronously and is provable.
The dealer should publish proofs of knowledge of each revocation
share, proving their creation and authenticity. Revocation issuers can
also use the dealer proofs to verify the validity of the shares
received without revealing them. Such revocation proofs will also be
useful when revocation issuers will reconstruct a revocation, since
they can refer to them to prove their shares are authentic without
revealing their content.

\section{Benchmarks}

We implemented the flows for credential issuance, presentation,
verification and revocation for lab tests using
Zenroom \footnote{Zenroom home: https://zenroom.org}, a secure
isolated execution environment implementing advanced cryptography
transformations. The reference implementation for this paper is
published on a public repository \footnote{SD-BLS github repository:
https://github.com/dyne/sd-bls}. All benchmarks were executed on a 6th
gen. Intel PC running tests on a single i7 3.40GHz core and making no
use of hardware acceleration.

\begin{figure}
    \centering
    \includegraphics[width=1\linewidth]{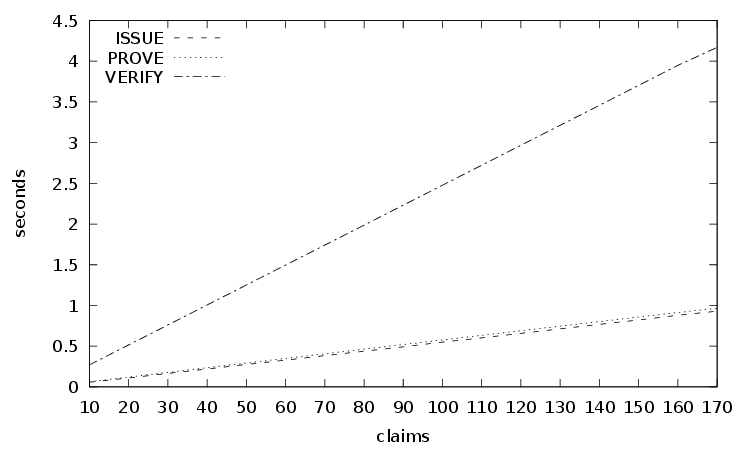}
    \caption{Speed comparison of issue and verify in SD-BLS}
    \label{fig:issueproveverify}
\end{figure}

Lab measurements on a growing number of claims show that issuance is
less computation heavy than verification, as shown in figure
\ref{fig:issueproveverify}.

As a handy reference, we provide a benchmark of the same operations
performed using the BBS+ in Zenroom on the same machine: results show
that SD-BLS is one order of magnitude slower than BBS+, whose
verification runs aproximately at the same speed as other operations.

\begin{figure}
    \centering
    \includegraphics[width=1\linewidth]{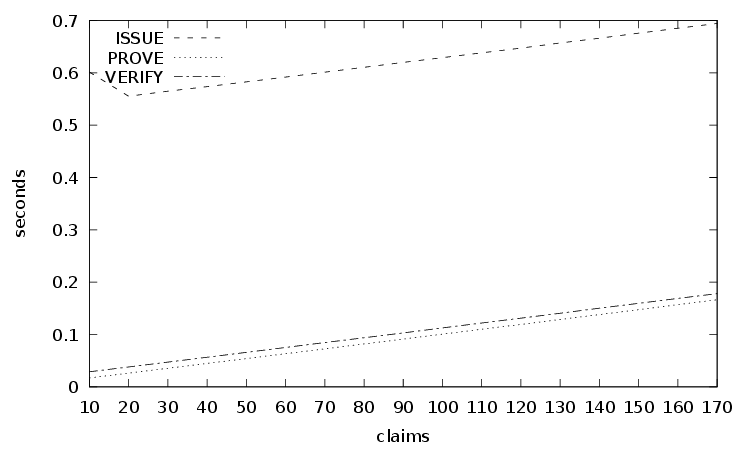}
    \caption{Speed comparison of issue and verify in BBS+}
    \label{fig:bbs_issueproveverify}
\end{figure}

Based on our benchmarks, the resulting data objects sizes in SD-BLS
are:
\begin{itemize}
    \item Signed claim: 177 Bytes
    \item Proof:  322 Bytes
    \item Revocation: 32 Bytes
\end{itemize}

\begin{figure}
    \centering
    \includegraphics[width=1\linewidth]{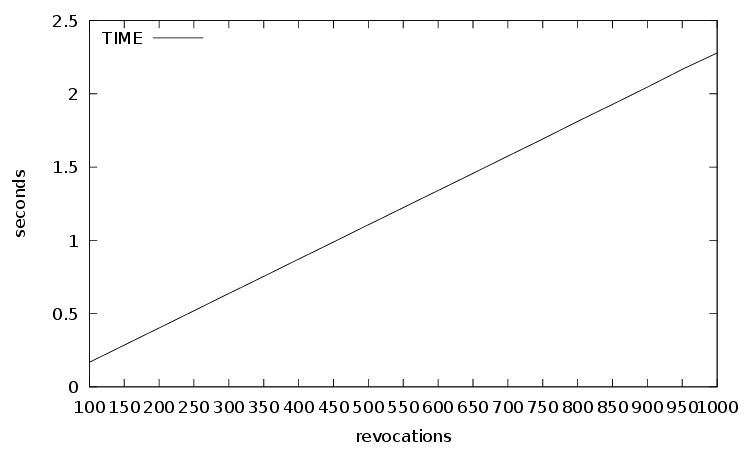}

    \caption{Speed of verification of a claim over multiple revocations}
    \label{fig:verifyrevocations}
\end{figure}

The computational cost of verifications grows linearly in presence of
a cryptographic revocations list. We assume revocations are published
as lists of secret revocation keys. Lab measurements of the time taken
by a single proof verification process to operate on a growing number
of revocations is shown in figure \ref{fig:verifyrevocations}.

The threshold operated for revocation issuers consists of a verifiable
secret sharing implementation supporting a configurable total and
quorum of peers. Our implementation shows very good performance on
reconstruction, which is the most speed-sensitive operation in
scenarios where responsive revocation process is required. The process
of reconstruction can be easily scaled for asynchronous consensus on a
micro-service swarm architecture and verified on blockchain.

\begin{figure}
    \centering
    \includegraphics[width=1\linewidth]{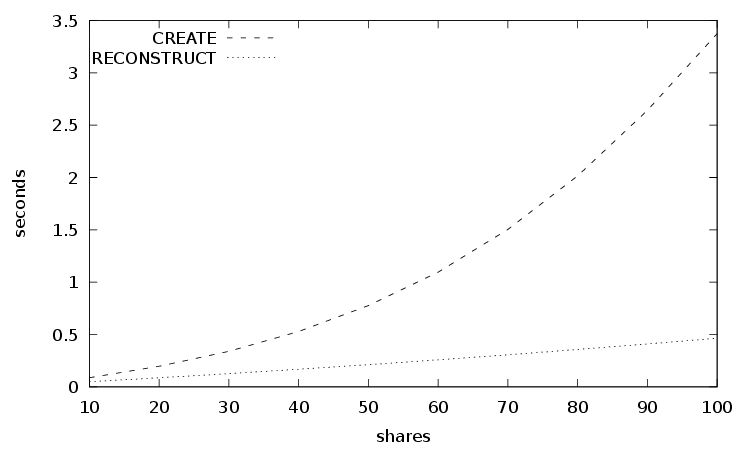}
    \caption{Speed of creation and reconstruction of shares among multiple peers}
    \label{fig:pvss}
\end{figure}


\section{Conclusion}
\subsection{Security considerations}

The revocations database is privacy and corruption sensitive (by our
previous definition of issuer corruption) and it should be securely
stored by each Issuer. This is mitigated by the adoption of threshold
revocation. When using threshold revocation there is still the need
for a \textit{revocation dealer} who has no access to private
informations, but may be dishonest and publish revocation keys
collected during the process of credential issuance.

BLS signatures and the proof system obtained with credentials are
considered secure by assuming the existence of random oracles
\cite{random-oracle}, together with the decisional Diffie-Hellman
Problem (DDH) \cite{DDH-problem}, the external Diffie-Hellman Problem
(XDH), and with the Lysyanskaya-Rivest-Sahai-Wol Problem (LRSW)
\cite{lrsw-assumption}, which are connected to the Discrete
Logarithm. The future growth of quantum-computing technologies may be
able to overcome the Discrete Logarithmic assumptions by qualitatively
different computational means and SD-BLS may then be vulnerable to
quantum-computing attacks. However this is speculative reasoning on
what we can expect from the future.

The SD-BLS implementation we are presenting in this paper is
demonstrated using the BLS12-381 curve \cite{bls381-12} also adopted
by ETH2.0. Debating the choice of BLS12-381 is beyond the scope of
this paper, but is worth mentioning that we can easily switch using
the BLS461 curve based on a 461 bit prime, hence upgrading our
implementation to 128 bit security \cite{updating-key-pairings}
against attacks looking for discrete logs on elliptic curves
\cite{discrete-log-attack}.

The $H$ component may be protected against brute-forcing attacks using
hash collisions to forge a valid credential for a different message. A
protection against this attack can be the adoption of a key-derivation
function like Argon2 \cite{argon2} on hash creation, which will add
computational costs to issuance and verification.

Anyone with knowledge of $H,r$ can try to guess $m$ by appending known
strings, which becomes trivial especially in case of boolean
credential strings (i.e. ``above18=true''). We recommend the
credential issuer appends a nonce like $m:nonce$ before hashing and
signing the message, which will be known by the holder and always
disclosed when disclosing $m$.

\subsection{Future development}

In this section we describe possibilities for expanding the algorithm
to cover further applications, which appear promising while requiring
further investigation.

\paragraph{Compatibility with EUDI-ARF}

EUDI-ARF dictates that the holder's secret key generation and
signatures must occur inside a trusted platform module (TPM). For
mobile devices, this limits the secret keys and signatures to those
offered, via proprietary APIs by the mobile OS, namely RSA (multiple
flavours) and ECDSA on the secp256r1 curve. Currently the TPMs APIs
supported by Android and iOS do not support BLS 12-381 key generation
or signature.

Client-side signatures in EUDI-ARF are mostly used in the
authentication process, specifically in the proof of possession
required by the OpenID4VCI\cite{OID4VCI} issuance flow, but not in the
verification.

Therefore, we can investigate the possibility to use SD-BLS to
implement a partially retro-compatible superset of EUDI-ARF, by
maintaining the current issuance and verification protocols and using
an extended SD-JWT format. Also we can explore useful integrations
with the European Blockchain Services Infrastructure (EBSI
\cite{ebsi}).

\paragraph{Signroom and DIDroom}

In SD-BLS both credential issuers and revocation issuers are in charge
of various interactive administrative operations, while the dealing of
revocation shares can be easily automated.

We plan to integrate SD-BLS in the free and open source software
``Signroom'', an application we developed in the context of the NGI
ASSURE grant, and ``DIDroom'' the dashboard connected to our
\emph{did:dyne} W3C DID domain.

\paragraph{Digital Product Passport}
Efforts in standardization of Digital Product Passport (DPP) are
ongoing in both the EU (Cirpass, BatteryPass, Trace4EU) and US
(DSCSA). An obstacle to adoption of DPP technologies is the reluctancy
of manufacturers to share information about their supply-chain,
knowing that the information would become publicly available and
immutable due to blockchain storage. While requiring further analysis
and investigation, a further development of the SD-BLS scheme could
allow creating DPPs built on the selective disclosure principles,
which may facilitate the adoption of the technology in the industry by
preserving the privacy of natural persons present in the DPP as REA
agents\cite{reflow}, while authenticating their contribution.

\paragraph{DAO Technologies}
The SD-BLS math is fully compatible with ETH2.0 and can be computed
inside an Ethereum VM. A verifier implemented in solidity can be a
building block for more advanced Distributed Autonomous Organizations
(DAO \cite{dao}) that want to authenticate peers using the selective
disclosure of verifiable credentials instead of a single key based
proof of possession.

\section{Acknowledgements}
This work has been funded by the EU in the framework of the NGI
TRUSTCHAIN project, grant No 101093274. We thank Puria Nafisi Azizi
and Matteo Cristino for their infatiguable work on Zenroom, Luca Di
Domenico for his help on the PVSS implementation, Giuseppe De Marco
for sharing findings from his pioneering journey and Simone Onofri for
his insights on threat models. We extend our thanks the anonymous
reviewers for their valuable advice and the De Cifris association for
facilitating connections over a large network of unique professionals.

\bibliographystyle{IEEEtran}
\bibliography{references}

\end{document}